\documentclass[sigconf]{acmart}
\AtBeginDocument{%
  \providecommand\BibTeX{{%
    \normalfont B\kern-0.5em{\scshape i\kern-0.25em b}\kern-0.8em\TeX}}}

\setcopyright{acmcopyright}
\copyrightyear{2024}
\acmYear{2024}
\setcopyright{acmlicensed}\acmConference[WWW '24 Companion]{Companion Proceedings of the ACM Web Conference 2024}{May 13--17, 2024}{Singapore, Singapore}
\acmBooktitle{Companion Proceedings of the ACM Web Conference 2024 (WWW '24 Companion), May 13--17, 2024, Singapore, Singapore}
\acmDOI{10.1145/3589335.3648319}
\acmISBN{979-8-4007-0172-6/24/05}

\usepackage{multirow}
\usepackage{marvosym}
\usepackage{diagbox}
\usepackage{graphicx}
\usepackage{balance}
\begin{document}


% pepnet 
% parameter and embedding personalized network for infusing with personalized prior information

\title{MetaSplit: Meta-Split Network for Limited-Stock Product Recommendation}

% in Large-scale Recommender System


\author{Wenhao Wu}
\authornote{Wenhao Wu and Jialiang Zhou contributed equally to this research. This work was done when Wenhao Wu was a research intern at Alibaba Group. Shuguang Han is the corresponding author.}
\orcid{0009-0009-9522-0703}
\email{wwh137@stu.xjtu.edu.cn}
\affiliation{%
  \institution{Xi’an Jiaotong University}
  \city{Xi'an}
  \country{China}
}

\author{Jialiang Zhou}
\authornotemark[1]
\email{zhoujialiang.zjl@alibaba-inc.com}
\affiliation{%
  \institution{Alibaba Group}
  \city{Hangzhou}
  \country{China}}

\author{Ailong He}
\email{along.hal@alibaba-inc.com}
\affiliation{%
  \institution{Alibaba Group}
  \city{Hangzhou}
  \country{China}}

\author{Shuguang Han\textsuperscript{\Letter}}
% \authornote{Corresponding Author}
\email{shuguang.sh@alibaba-inc.com}
\affiliation{%
  \institution{Alibaba Group}
  \city{Hangzhou}
  \country{China}}

\author{Jufeng Chen}
\email{jufeng.cjf@alibaba-inc.com}
\affiliation{%
  \institution{Alibaba Group}
  \city{Hangzhou}
  \country{China}}

\author{Bo Zheng}
\email{bozheng@alibaba-inc.com}
\affiliation{%
  \institution{Alibaba Group}
  \city{Hangzhou}
  \country{China}}


\renewcommand{\shortauthors}{Wenhao Wu et al.}
\begin{abstract}
Compared to business-to-consumer (B2C) e-commerce systems, consumer-to-consumer (C2C) e-commerce platforms usually encounter the limited-stock problem, that is, a product can only be sold one time in a C2C system. This poses several unique challenges for click-through rate (CTR) prediction. Due to limited user interactions for each product (i.e. item), the corresponding item embedding in the CTR model may not easily converge. This makes the conventional sequence modeling based approaches cannot effectively utilize user history information since historical user behaviors contain a mixture of items with different volume of stocks. Particularly, the attention mechanism in a sequence model tends to assign higher score to products with more accumulated user interactions, making limited-stock products being ignored and contribute less to the final output. To this end, we propose the Meta-Split Network (MSNet) to split user history sequence regarding to the volume of stock for each product, and adopt differentiated modeling approaches for different sequences. As for the limited-stock products, a meta-learning approach is applied to address the problem of inconvergence, which is achieved by designing meta scaling and shifting networks with ID and side information. In addition, traditional approach can hardly update item embedding once the product is consumed. Thereby, we propose an auxiliary loss that makes the parameters updatable even when the product is no longer in distribution. To the best of our knowledge, this is the first solution addressing the recommendation of limited-stock product. Experimental results on the production dataset and online A/B testing demonstrate the effectiveness of our proposed method.

\end{abstract}






\maketitle

\section{Introduction}
% 1 大背景 
% 2 小背景
% 3 小领域研究现状 -> 研究现状不足 -> 本文研究目标
% 4 本课题的重要性 和 独创性
% 5 研究问题 -> 解决办法
% 6 研究贡献
% 7 小结

% （1）Introduction部分要提供作者研究的背景知识、科学问题及该问题的重要性；
% （2）给该问题一个小的总结（综述）；
% （3）提出该问题是否存在争议；
% （4）是否完成了或部分解释了提出的科学问题。

% -----------------------------------------
% 大领域背景（RS）+小领域背景(CTR)+研究现状(CTR Prediction)

\begin{figure}[tbp]
    \centerline{\includegraphics[width=6.5cm]{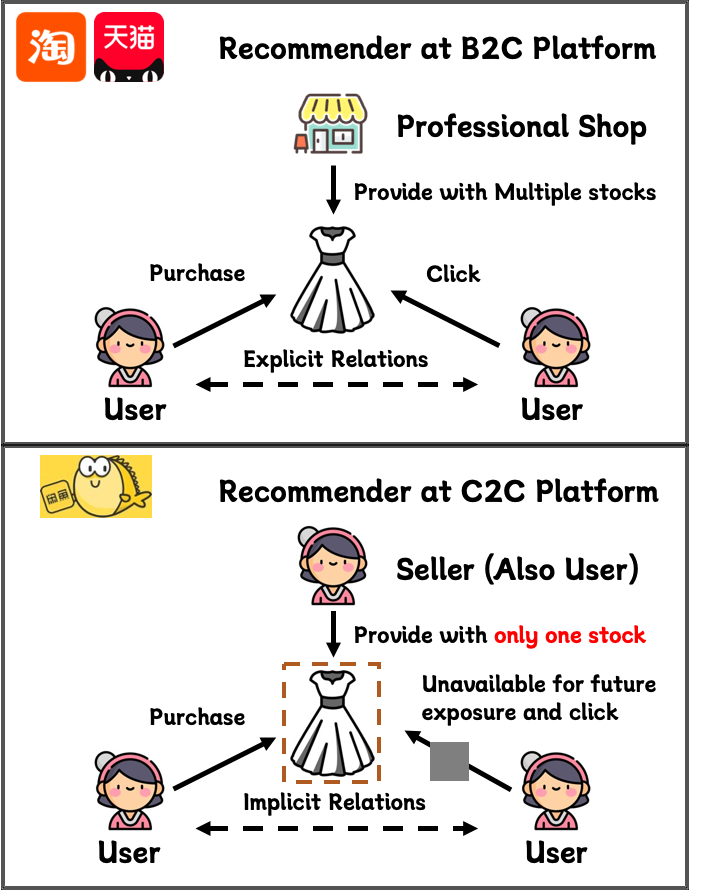}}
    \caption{An illustrative example highlighting the distinctions between B2C platforms and C2C platforms.}
    \label{fig:c2c_example}
    % \vspace{-0.4cm}
\end{figure}

% \begin{table}[!t]
%     % \renewcommand\arraystretch{0.9}
%     % \small
%     % \setlength{\abovecaptionskip}{0.05cm}
%     % \captionsetup{font={small}}
%     \centering
%     \begin{tabular}{ccc}
%     \toprule
%     & \textbf{Type B Item} & \textbf{Type C Item}\\
%     \midrule
%     \textbf{Seller} & Professional Seller & Normal Consumer \\
%     \textbf{Typical Example} & New Wallet & Used Wallet \\
%     \textbf{Item Visual Quality} & High & Inconsistent \\
%     \textbf{Daily Exposures} & High ($>$500) & Low ($<$10) \\
%     \textbf{After Selling} & Re-listing & Off the market \\
%     \bottomrule
%     \end{tabular}
%     \caption{
%         Different characteristics between type-business items and type-costumer items.
%     }
%     \label{tab:char_summary}
%     % \vspace{-0.4cm}
% \end{table}

In today's digital era, recommendation systems are crucial in various domains like e-commerce, streaming services, and social media platforms~\cite{cvr1, din,youtubednn,tencent_www}. These systems utilize user data to learn user interests,  estimate preferences from collected user interactions, and provide personalized suggestions~\cite{gift,cvr2,cvr3,cvr_xianyu,zhang2022keep}, thereby enhancing user experience and driving sales.  An accurate click-through rate (CTR) prediction model is vital in achieving this goal. 

% add a lot of reference here
In recent years~\cite{shallow1,mmoe,pnn}, industrial click-through rate prediction models have shifted from traditional shallow models~\cite{shallow2,shallow3} to deep learning models~\cite{star,ple,wu2022adversarial}. The deep CTR prediction model generally follows an Embedding and MLP architecture~\cite{wideanddeep}. The embedding layer transforms each discrete feature from raw input into a low-dimensional vector. The deep model can take various forms, such as models that learn high-order interactions or take into account historical behavior sequences. By utilizing deep models, these systems can effectively capture complex patterns and relationships in user data, leading to more accurate predictions of user engagement. These methods have demonstrated state-of-the-art performance across a wide range of tasks~\cite{s3,m6}.

% 我们要研究的问题
The interaction between products and consumers can go beyond the traditional Business-to-Consumer (B2C) model, and the Consumer-to-Consumer (C2C) model has emerged in recent years. While B2C platforms involve transactions between businesses and individual consumers, C2C platforms facilitate transactions among individual consumers themselves. As China's largest online C2C platform, Xianyu allows both individuals and small professional shop owners to sell their products. This paper conducts all its experiments within the Xianyu platform, leveraging its unique C2C dynamics for insightful analysis.
% 详细介绍b2c，c2c
% B2C平台的商品由专业商家提供，有很多库存，C2C平台上的商品由用户发布，库存有限，库存量有限。请注意，尽管C2C平台主要由用户发布商品，但也存在由专业商家提供的深库存商品。这些商品通常是某些商家专门选择在C2C平台上销售，以满足特定的需求或市场。这样的商品可能具有更高的库存量和专业性，与普通用户发布的商品有所区别

The products on B2C platforms are mainly from professional merchants and these products always have a wide range of stock available. These merchants operate as businesses and specialize in selling products directly to consumers. Meanwhile, the products on C2C platforms are mainly posted by individual users, which often have limited stock availability (e.g., a second-hand baby stroller with only one stock). 
The main characteristic of limited-stock products is that once sold, they are no longer available on the platform. The distinction between B2C and C2C can be simply visualized as the example shown in Fig. \ref{fig:c2c_example}. 
However, it is important to note that although C2C platforms mainly consist of limited-stock products, there are also products with multiple-stocks provided by professional merchants.
% 细致分析
% 在C2C场景上，浅库存商品占比很高，累计反馈很少，长尾问题很严重，曝光比较少，冷启比较麻烦，存在冷启动问题

There are two major challenges arising from the recommendation system on C2C platform.
\begin{itemize}
    \item 
    % This will result in the spares interactions between limited-stock items and consumers, causing slower convergence rates or potential failure of the corresponding item embedding to converge. 
    The existence of limited-stock products makes the interaction between items and consumers much more sparse, which usually causes slower convergence rates or even failure in learning unique item ID embedding during model training. It has been widely known in the industry that a well-trained ID embedding can largely improve the recommendation performance~\cite{one_epoch}. Consequently, the problem of limited-stock products poses a challenge in accurately estimating the click-through rate (CTR), and subsequently affects the precision of CTR prediction, especially for products with limited stocks availability.
    
    \item The coexistence of these two types of items in the user historical behavior sequence complicates the user interest modeling process. In particular, attention mechanisms, commonly used in sequence modeling, may overlook the limited-stock product due to the behavior sparsity. Compared to limited-stock items, multiple-stock items usually have much more user interactions. This can cause the result recommendation to be biased towards multi-stock items. Thus, greatly affecting the overall modeling effectiveness. 
    %Currently, in C2C platforms, conventional sequence modeling approaches struggle to effectively leverage user history information due to limited user interactions for each product. Notably, attention mechanisms in sequence models tend to assign higher scores to products with greater accumulated user interactions, resulting in limited-stock products being overlooked and contributing less to the final output.
\end{itemize}
% A major challenge arising from the recommendation on C2C platform is the issue of limited stock, where certain products with limited stocks can only be sold once, and once they are sold, they are no longer available for purchase and click, as an example shown in Fig. \ref{fig:example}.
% Currently, in C2C platforms, conventional sequence modeling approaches struggle to effectively leverage user history information due to limited user interactions for each product. This issue makes it difficult for the corresponding item embedding in CTR models to converge. Notably, attention mechanisms in sequence models tend to assign higher scores to products with greater accumulated user interactions, resulting in limited-stock products being overlooked and contributing less to the final output.
% -----------------------------
% Recommendation for C2C platform has been neglected (neglect the fact that different item have different shangpin charterlistic, and different stocks) by the current research, but the problem is shared across item-cold start problem, long-tail problem, across both b2c and c2c platform,   they xxx
% The challenge for CTR prediction considering the limited stock product on C2C platform has been neglected by current research.
A common solution to address the above problems is to treat the limited-stock products as new items using item cold-start methods. However, this cannot fully address this issue, and the different volumes of product is being neglected.
% 常见的解法，按照item

To address these limitations, this paper proposes a novel approach called the Meta-Split Network (MSNet), which splits the user history sequence based on the volume of stock for each product and employs different modeling methods for different sequences. For multi-stock products, we apply the traditional Deep Interest Model~\cite{din}; and for limited-stock products, we introduce a meta-learning approach to handle the problem of model convergence. Specifically, we design meta-scaling and shifting networks with both ID and side information (such as category) to better utilize item information.

In addition, the traditional approach can hardly update item embedding once the product is consumed. Thereby, we propose an auxiliary loss to facilitate parameter updates even when the product is no longer in distribution. By incorporating this auxiliary loss, we can continue to refine and adjust the embedding even after the product has been consumed. This approach allows for learning a better item representation and helps a better understanding of products even if they are longer available for recommendation. It enables the model to capture more accurate and up-to-date information about the products, resulting in an improved CTR prediction performance.

To evaluate the effectiveness of our proposed method, we conducted experiments on a production dataset and performed online A/B testing. The results demonstrate the superiority of our approach over existing methods in accurately predicting CTR for overall products and limited-stock products items. We believe that this is the first solution to effectively address the recommendation of limited-stock items in e-commerce settings, opening up new possibilities for further improvement.

% -----------------------------
% 贡献
To summarize, the main contributions of our work are as follows:
\begin{itemize}
    \item To the best of our knowledge, this is the first solution to deal with the recommendation of limited-stock products in C2C e-commerce platforms. We reveal that there are key components for addressing this issue,  non-convergence embedding representation for limited-stock products and the mixed user behavior sequences containing both multiple-stocks products and limited-stock products. We believe this can open up new possibilities for further improvement.
    
    \item We propose a novel method named Meta-Split Network(MSNet) to address the challenges faced by consumer-to-consumer (C2C) e-commerce platforms, specifically in the context of limited-stock products. MSNet 
    adopt differentiated modeling approaches for user history sequences based on the volume of product stock. For limited-stock products, a meta-learning approach is employed to overcome issues of in-convergence, utilizing meta scaling and shifting networks with ID and side information. Additionally, the paper introduces an auxiliary loss to update item embedding even after the product is consumed.
    
    \item Experimental results on industrial production datasets and online A/B testing demonstrate the superiority of MSNet over existing methods in accurately predicting CTR for overall products and limited-stock products items.
    %to demonstrate the effectiveness of MSNet through offline test and online A/B testing. 
\end{itemize}

\section{Related Work}
This section briefly introduces the progress in the field of cold-start item recommendation and meta-learning for recommendation.

% In this section, we will introduce the related work from three aspects: Cold-start Recommendation, Meta Learning, Sequential Recommendation

\subsection{Cold-Start Recommendation}
% cold start有两个阶段，先有专门的冷启动链路，后有和有一些曝光后进入主链路和别的商品一起竞争。
% 加信息，user，item，side，更新的更快，增强实时性，但这些只是信息量和工程建设base的提升
% 第二个阶段，有两种方法，从样本的角度进行数据增强，从模型的角度进行定制，有dropoutnet，poso，定制一个差异化的网络
% 除此以外，一些关于长尾问题的研究也与item cold start类似，都是为了更好的预估冷门低曝光商品的CTR
%\textbf{Cold-Start Item Recommendation}:

Recommendations for new items, or the cold-start item recommendation problem, can be challenging due to the limited interactions between users and items~\cite{wu2022adversarial}. Cold-start item recommendation is typically divided into two stages by the amount of user interactions. 

In the first stage, the new items have just entered into the system and thus received very limited exposure. At this moment, any existing recommendation models may fail because of the minimal user interactions. To avoid direct competition with existing items, industrial recommendation systems may offer a separate channel only for new items. In this case, high-quality new items only compete with each other, and they can warm up and enter the next stage after obtaining a certain level of interaction. After that, one can calculate simple real-time statistical information or incorporate a limited amount of user and item features for recommendation. The common approaches ~\cite{cold_cv,cold_context,cold_cross_domain,cold_graph,cold_poi} utilize generalization features or explore information across different domains and modalities. The most effective algorithms, of course, are usually the ones that involve introducing additional user-item interaction data from other domains~\cite{zhang2022keep,rec4ad}.

In the second stage, new items may receive a few interactions from the separate channel, And now, they compete directly with the existing items. 
ClassBalance~\cite{class_re} and LogQ~\cite{logq} attempted to enhance the importance of new items during the training process by adjusting the weights of the samples or their corresponding network parameters. 
DropoutNet~\cite{dropoutnet} and POSO~\cite{poso} enhanced the expressive ability and robustness of models for new items by designing specialized networks tailored to their characteristics. There are indeed many other research studies on this topic, we do not list them all here due to the content limit.

\subsection{Meta Learning for Recommendation}
% learning to learn
% melu，mwuf，mamo，让冷启商品去适配在线网络
Meta learning enables new tasks to be learned on a small number of samples by learning the general or auxiliary knowledge. 
% Meta learning methods in recommendation have gained significant attention in recent years. These methods leverage the insights obtained from multiple recommendation tasks to improve the performance of individual recommendation models. 
Methods like MetaEmb \cite{meta_emb} and MWUF \cite{mwuf} leverage additional information from items and users to generate or adjust item ID embedding. 
Techniques such as model-agnostic meta learning (MAML) \cite{MAML} and metric-based meta learning have been applied to learn transferable representations or optimize the learning process across different recommendation scenarios. 
% In contrast to previous works, our focus lies in applying meta learning to address the challenges of cold-start item recommendation and provide more accurate and personalized recommendations for new users or items.
\section{Preliminaries}

\begin{figure}[tbp]
    \centerline{\includegraphics[width=8cm]{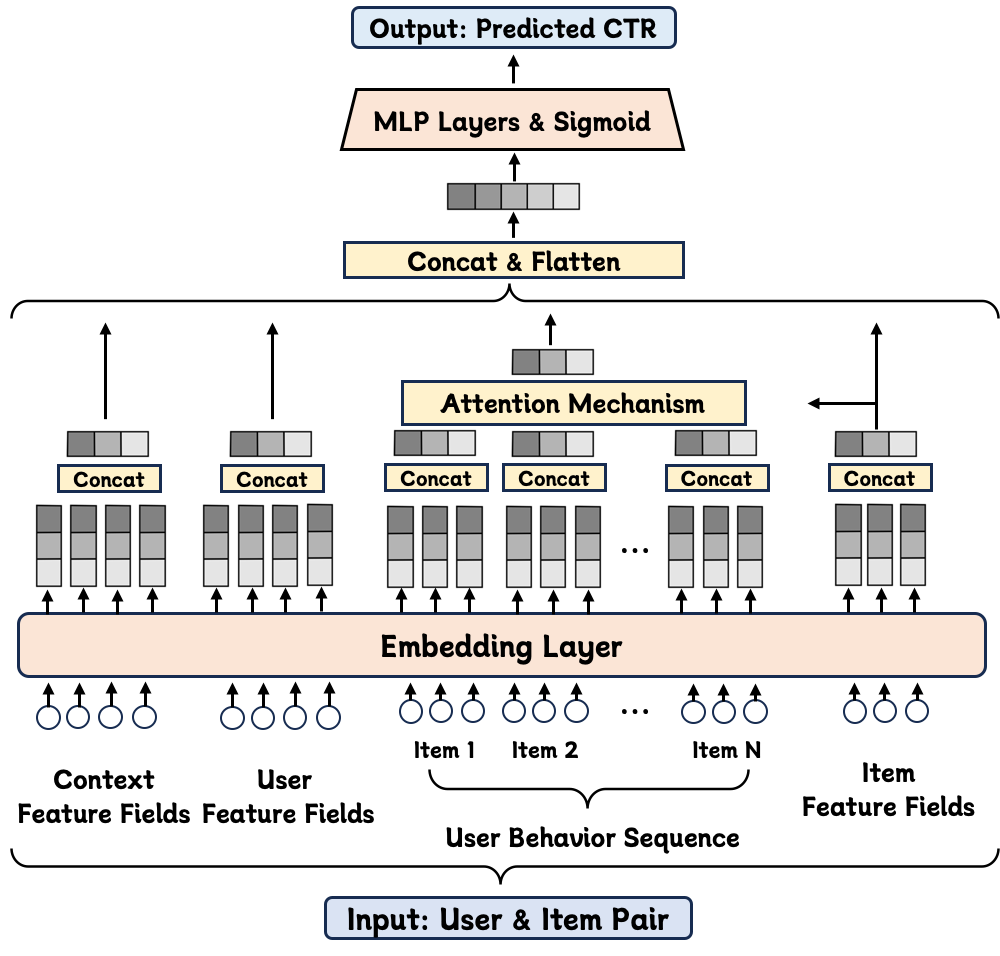}}
    \caption{An illustration of Embedding and MLP structure with sequntial information modeling for the deep CTR prediction model}
    \label{fig:din_model}
    % \vspace{-0.4cm}
\end{figure}

In this section, we formulate the problem and briefly introduce the existing production CTR prediction model on the large-scale recommender system of Xianyu.

% \textit{\textbf{Problem Formulation}}: 
\subsection{Problem Formulation}
Given a dataset $\mathcal{D}=\{ \mathbf{x}, y \}^N,(\mathbf{x}, y)$ marks a sample and $N$ is the number of samples, where $\mathbf{x}$ denotes the high dimensional feature vector consisting of multi-fields (e.g., user and item field), and $y$ is the binary label with $y = 1$ indicating the sample is clicked. Our task is to accurately predict the probability of CTR $p_{ctr} = p(y=1|x)$ for the testing sample $x$.

\subsection{Production CTR Prediction Model}

We select Deep Interest Network \cite{din} as our base model due to its online efficiency and effectiveness, which follows the conventional Embedding and MLP paradigm and utilizes an attention mechanism to model user behavior sequences, as depicted in Figure~\ref{fig:din_model}.

\textit{\textbf{Embedding Layer}}: 
The inputs are composed of non-sequential features (e.g., user ID) and sequential features (e.g., user’s history clicked items). The embedding layer is employed to convert each discrete feature from the raw input into a vector of lower dimensions, by using an embedding look-up table. The embedding of non-sequential features is simply concatenated, whereas for the embedding of sequential features, a sequence information modeling module is used to assemble them into a fixed-size representation.

\textit{\textbf{Sequence Information Modeling}}: 
Deep Interest Network(DIN) utilizes a local attention mechanism to dynamically capture the user's interests based on the similarity between their historical clicked items and the target item. 
This mechanism allows for the acquisition of a personalized representation of the user's interests since it enables weight pooling of the information in the sequence of varying length.
Additionally, DIN could be further optimized by leveraging the target attention mechanism equations~\cite{BST,transformer,pssa} to replace the original local attention mechanism.

In the existing Multi-head Target-attention (MHTA) implementation, the target item $Item_t$ is consider as query $(K)$ and the history click sequence $\boldsymbol{S}_u$ is considered both as keys $(K)$ and values $(V)$, where $\boldsymbol{S}_u=\left\{Item_1, Item_2, \ldots, Item_H\right\}$ is the set of embedding vectors of items in the user behaviors with length of $H$. 
% Then, they are produced by a linear projection layer. $Q =W^Q Item_t$, $K=W^K X^{\text {seq }}$, and $V=W^V X^{\text {seq }}$, where ($W^Q$, $W^K$ and $W^V$ are trainable matrices)

Specifically, the output of MHTA can be formalized as follows :

% \begin{equation}
% \begin{gathered}
% \boldsymbol{h}_u=\text { TargetAttention }\left(\boldsymbol{h}_t \boldsymbol{W}^Q, \boldsymbol{H}_u \boldsymbol{W}^K, \boldsymbol{H}_u \boldsymbol{W}^V\right) \\
% \end{gathered}
% \end{equation}
\begin{equation}
\text { TargetAttention }(Q, K, V)=\operatorname{softmax}\left(\frac{Q K^{\top}}{\sqrt{d}}\right) V
\end{equation}
where $Q =W^Q Item_t$, $K=W^K X^{\text {seq }}$, and $V=W^V X^{\text {seq }}$, the linear projections matrices $W^Q \in \mathbb{R}^{d \times d}, W^K \in \mathbb{R}^{d \times d}, W^V \in$ $\mathbb{R}^{d \times d}$ are learn-able parameters and $d$ stands for the dimension of hidden space. The temperature $\sqrt{d}$ is introduced to produce a softer attention distribution for avoiding extremely small gradients.

Finally, all non-sequential embedding and transformed user sequential embedding is concatenated with other continuous features together to generate the overall embedding and pass through a deep network to get the final prediction.

\textit{\textbf{Loss}}: 
The objective function used in DIN and our proposed MSNet is both the negative log-likelihood function defined as:
\begin{equation}
L=-\frac{1}{N} \sum_{(x, y) \in \mathcal{D}}(y \log f(x)+(1-y) \log (1-f(x)))
\end{equation}
where $D$ is the training set, each sample $x$ is associated with a ground-truth label $y$. The output of the model, denoted as $f(x)$, represents the predicted probability of sample $x$ being clicked.

\section{Methodology}\label{sec:method}

%   \includegraphics[width=\textwidth]{acm-jdslogo.png}

\begin{figure*}[tbp]
    \centerline{\includegraphics[width=18cm]{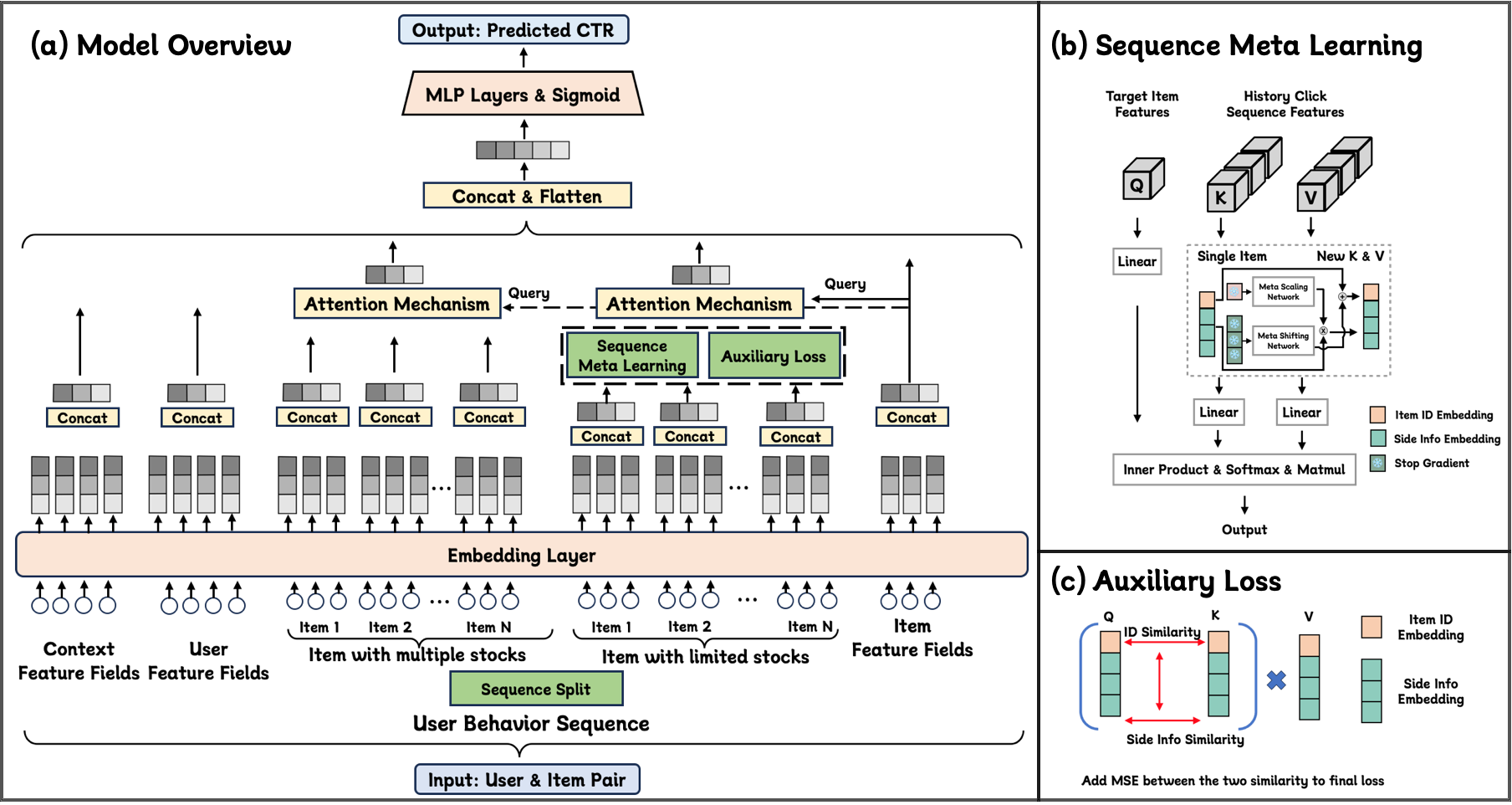}}
    \caption{The schematic framework of MSNet can be broadly divided into three components: the sequence split module, the sequence meta-learning module, and the auxiliary loss. Figure (a) provides an overview of the model, with the sequence split module depicted at the bottom. Figure (b) illustrates the sequence meta-learning module, while Figure (c) offers detailed insights into the auxiliary loss.}
    \label{fig:model_overview}
    % \vspace{-0.4cm}
\end{figure*}

\begin{table}[!t]
    % \renewcommand\arraystretch{0.9}
    % \small
    % \setlength{\abovecaptionskip}{0.05cm}
    % \captionsetup{font={small}}
    \centering
    \begin{tabular}{ccc}
    \toprule
    {\diagbox{\textbf{Item}}{\textbf{Sequence}}}& \textbf{M/s Product} & \textbf{L/s Product}\\
    \midrule
    \textbf{M/s Product} & 0.35 & 0.17 \\
    \textbf{L/s Product} & 0.09 & 0.07 \\
    \bottomrule
    \end{tabular}
    \caption{
        Analysis of average attention scores (before softmax function) in DIN Module, where "M/s" is short for "multiple-stocks" and "L/s" is short for "limited-stock".
    }
    \label{tab:attention_score}
    % \vspace{-0.4cm}
\end{table}

% \begin{table}[!t]
%     % \renewcommand\arraystretch{0.9}
%     % \small
%     % \setlength{\abovecaptionskip}{0.05cm}
%     % \captionsetup{font={small}}
%     \centering
%     \begin{tabular}{ccc}
%     \toprule
%     {\diagbox{\textbf{Item}}{\textbf{Sequence}}}& \textbf{M/s Product} & \textbf{L/s Product}\\
%     \midrule
%     \textbf{M/s Product} & 0.35 & 0.17 \\
%     \textbf{L/s Product} & 0.09 & 0.07 \\
%     \bottomrule
%     \end{tabular}
%     \caption{
%         Analysis of norm of ID Embedding for different type of product, where M/s and L/s remain the same meaning.
%     }
%     \label{tab:attention_score}
%     % \vspace{-0.4cm}
% \end{table}

This chapter outlines the comprehensive design of our method, including the architecture overview of our proposed method, Meta-Split Network (MSNet), and the detailed structure.

\subsection{Architecture Overview}
Ignoring the fact that limited-stock product have insufficient embedding representation and not distinguish limited-stock product in user sequence will leads to inferior model performance.
To this end, we propose the Meta Split Network (MSNet) for CTR prediction to better utilize limited-stock items by adopting differentiated modeling approaches for user history behavior sequences.  As shown in Figure \ref{fig:model_overview}, MSNet consists of three main components: (1) the sequence split module (2) the sequence meta learning network (3) the auxiliary loss.

MSNet first split user behavior history sequence based on the volume of product stock.
For multi-stock products, we apply the traditional Deep Interest Model, and for limited-stock products, we introduce a meta-learning approach to handle the problem of model convergence. The meta learning module involves employing meta scaling and shifting networks to enhance ID and side information mutually. Furthermore, MSNet use an auxiliary loss that updates item embedding even after the product has been consumed. This ensures that the model continues to learn and adapt even when products are no longer available. Afterwards, the attention mechanism receives the enhanced user sequential embedding and process them into transformed user behavior interest.  The user behavior interest are further concatenate with the non-sequential features and passed to  the deep network for CTR prediction.

% % splits the sequence based on the volume of stock and models a user interest vector for each sequence. The deep inventory product sequence is modeled in the same way as the original DIN method, while the modeling of the shallow inventory product sequence have some differences. Finally, the sequences are concatenated with the non-sequential features. The final outputs of the embedding layer are concatenated and passed to the DNN network for prediction.

% MSNet is a model that takes into account the volume of product stock when modeling user history sequences. To handle products with limited stock, the model uses a meta-learning approach to address convergence issues. This involves employing meta scaling and shifting networks that leverage ID and side information.

% Furthermore, the paper presents an auxiliary loss that updates item embedding even after the product has been consumed. This ensures that the model continues to learn and adapt even when products are no longer available.

% In summary, MSNet utilizes differentiated modeling approaches for user history sequences based on product stock volume. It overcomes in-convergence issues for limited-stock products using meta-learning techniques and incorporates an auxiliary loss to update item embedding beyond product consumption.
% % It should be noted that MSNet is not dependent on any specific DNN model and can be easily generalized to state-of-the-art (SOTA) methods.
\subsection{Detail Structures}
In this section, we will introduce the components of our proposed method in detail.

\subsubsection{\textbf{Sequence Split Module}}
The user behavior sequence includes the items they have clicked with, including item ID and side information like category and genre. 
Concretely, representation of a single item and the whole sequence is given as:
The representation of a single item and the whole sequence can be written in the following way:
\begin{equation}
    \boldsymbol{E}_{\text {Item}}=Concat [\boldsymbol E\left(\mathcal{F}_{ID}\right), \boldsymbol E\left(\mathcal{F}_{Side}\right)]
\end{equation}

\begin{equation}
    \boldsymbol{E}_{\text {Seq}}=
    Concat [\boldsymbol{E}_{\text {Item\_1}}, \boldsymbol{E}_{\text {Item\_2}}, \ldots
    \boldsymbol{E}_{\text {Item\_k}}]
\end{equation}
where $\boldsymbol{E}_{\text {Item}},E\left(\mathcal{F}_{ID}\right),E\left(\mathcal{F}_{Side}\right),\boldsymbol{E}_{\text {Seq}}$ means the embedding for item, id feature, side feature and sequence feature respectively.

We split the user behavior sequence based on item type (deep stock or limited stock) and this will further facilitates the subsequent differential modeling of the two sequences to prevent high-frequency signals from overriding low-frequency signals.
where $\boldsymbol{E}_{\text {B\_Seq }}, \boldsymbol{E}_{\text {C\_Seq }}$ means the embedding for b seq and c seq respectively.
During the actual execution, this sequence split process can be achieved in the training samples construction process, or by using a mask mechanism during training.

Our threshold for distinguishing between multiple-stocks and limited-stock items is set at one item: products with a volume greater than one are categorized as multiple-stocks, and a product with only one item is identified as limited-stock.
% To address this issue, we first split the user's historical sequence of clicked items based on their types. Then, for the sequence of C-type items, we introduce a Meta Network module. Since the ID embedding signal of C-type items is relatively weak, simply isolating and aggregating them as input into the network still results in a weak signal that cannot effectively impact feature interactions and final predictions. To overcome this, we generate a meta ID based on the side information of C-type items, such as category and category information, and append it to the original ID of C-type items.

\subsubsection{\textbf{Sequence Meta Learning Module}}

Inspired by the concept of MWUF~\cite{mwuf}, which implements a meta scaling and shifting network to generate embedding representation for new items through existing embedding representation of old items, we propose sequence meta learning module. 
We have made modifications to its usage and applied it in a sequential manner, while maintaining the names of the two models.

We begin with the equation of attention mechanism, the equation is listed below:
\begin{equation}
    User Interest = att (Q * K) * V
\end{equation}
where Q is the target to match others, K is the Key to be matched, and V is the value to be multiplied after matching. In the context of CTR prediction, Q represents the item that needs to be inferred, while both K and V represent the user behavior sequence.

As we can see, Q and K are used for similarity matching, while V is utilized for the final output. Our objective is to achieve a more equitable matching between Q and K, where the side information with higher generalization capability plays a more vital role, especially when the ID embedding are not converged and lack sufficient training.
Simultaneously, we aim to enhance the information contained in the returned V. To accomplish this, we scale the side information of Q and K based on the sequence of Item IDs and generate a meta ID using the side information from the product sequence. Subsequently, this meta ID is employed to shift the information in V.

\textbf{Meta Scaling Network}:
We use ID feature to generate weight for each feature field of side information.
Specifically the scaling weight is generate by the following equation
\begin{equation}
\boldsymbol{\delta}_{\text {weight}} = DNN\left( \oslash\boldsymbol{E}\left(\mathcal{S}_{ID}\right) \right)
\end{equation}
where  $\boldsymbol{E}\left(\mathcal{S}_{ID}\right)$ means the ID embedding and $\oslash$ means stop gradient.

we scale the side feature embedding accordingly by a field-wise product manner
\begin{equation}
Re\_E\left(\mathcal{S}_{Side}\right) = \boldsymbol{\delta}_{\text {weight }} \otimes E\left(\mathcal{S}_{Side}\right)
\end{equation}

\textbf{Meta Shifting Network}:
By enhancing the ID information, we are able to better capture the unique characteristics and features of each limited-stock item.
Specifically the scaling weight is generate by the following equation
\begin{equation}
\boldsymbol{\delta}_{\text {Meta ID }} = DNN\left( \oslash\boldsymbol{E}\left(\mathcal{S}_{Side}\right) \right)
\end{equation}
where $\boldsymbol{E}\left(\mathcal{S}_{Side}\right)$ is the side information embedding of the sequence feature.
We want to make more use of meta ID information when the original ID is relatively weak, and less use of meta ID information when the original ID is relatively strong. Thus we add the meta ID embedding with the origin ID embedding w.r.t the norm ratio
\begin{equation}
Re\_E\left(\mathcal{S}_{ID}\right) = v \odot \boldsymbol{\delta}_{\text {Meta ID }}  \oplus (1-v) \odot \boldsymbol{E}\left(\mathcal{S}_{ID}\right)
\end{equation}
\begin{equation}
v = \frac{Norm(\boldsymbol{\delta}_{\text {Meta ID }})}{Norm(\boldsymbol{\delta}_{\text {Meta ID }}) + Norm(\boldsymbol{E}\left(\mathcal{S}_{ID}\right))}
\end{equation}
where $v$ is the embedding norm ratio for the $\boldsymbol{E}\left(\mathcal{S}_{ID}\right)$.

The final representation of a item is:
\begin{equation}
Final\_E\left(\mathcal{K}\right) =  Concat [E\left(\mathcal{S}_{ID}\right), Re\_E\left(\mathcal{S}_{Side}\right)]
\end{equation}
\begin{equation}
Final\_E\left(\mathcal{V}\right) =  Concat [Re\_E\left(\mathcal{S}_{ID}\right), E\left(\mathcal{S}_{Side}\right)]
\end{equation}

The scaling network aims to identify the importance of side information using ID information, while the shifting network aims to enhance the ID information using side information. Therefore, both of these networks can be classified as meta-learning networks.

\subsubsection{\textbf{Auxiliary Loss}}
% Q和K的相似度越小，就会导致传回V的梯度越小，导致梯度更新的幅度更小，导致越来越不相似

In the original attention mechanism, the items in the sequence are aggregated into a single vector through weighted pooling and passed into the DNN network. For a single item in the sequence, the gradient returned during back-propagation depend on the values passed during aggregation. As a result, items that contribute more significantly will have larger gradient updates, while items with lower contributions will have smaller gradient updates. This can lead to items in the sequence that are ignored due to insufficient training of their item id embeddings being continuously ignored and not receiving significant updates.

The introduction of an auxiliary loss is based on the assumption that items with similar categories should have similar ID embeddings. To implement this, we first calculate the cosine similarity between the sequence of items and the target item's category, as well as the cosine similarity between the sequence of items and the target item's ID. We then compute the mean squared error (MSE) between these two similarities and incorporate it into the final loss.

\begin{equation}
    Side\_sim = (cos\_sim(T_{Side}, S_{Side}))
\end{equation}

\begin{equation}
    ID\_sim = (cos\_sim(T_{Id}, S_{Id}))
\end{equation}

% \begin{equation}
%     Loss_{aux} = MSE(Side\_sim, ID\_sim)
% \end{equation}

To facilitate the updating of item id embedding for limited-stock items in the sequence, we can prevent the already trained side information embedding from being misleading. One approach is to apply the stop gradient operation to the similarity of the side information. This ensures that the gradients do not flow back to the side information embedding during the training process, allowing the focus to be primarily on updating the item id embedding.

\begin{equation}
    Loss_{aux} = MSE(\oslash Side\_sim, ID\_sim))
\end{equation}

\begin{equation}
    Loss = Loss_{CE} + \alpha * Loss_{aux}
\end{equation}

% \subsection{Non-sequential Information Embedding boost}
% For the non-sequential information part, We made some improvements based on POSO, PPNET, and Gatenetwork. 
% We will select some features with high correlation to the final prediction through feature analysis and freeze the gradient input into the Gate. At the same time, we will keep these features in the input of the DNN and multiply them field-wise with the values output by the Gate network to the original DNN input layer.

% \begin{equation}
%     \boldsymbol{E}_{\text {DNN Input }}=E\left(\mathcal{F}_{1}\right) \oplus E\left(\mathcal{F}_{2}\right) \oplus E\left(\mathcal{F}_{3}\right)
% \end{equation}

% \begin{equation}
% \boldsymbol{\delta}_{\text {weight }} = DNN\left( \oslash\left(\mathcal{F}_{Selected}\right) \right)
% \end{equation}

% \begin{equation}
% Re\_E\left(\mathcal{F}_{DNN Input}\right) = \boldsymbol{\delta}_{\text {weight }} \otimes E_{\text {DNN Input }}
% \end{equation}

% And finally, non-sequential embedding and sequential embedding are concatenation to pass through the DNN network.

Please note that our model mainly focuses on the Embedding module in the input layer of the DNN, which can be applied to various DNN networks and sequence information modeling method using the attention mechanism.
\section{Experiments}

\begin{table*}[t]
\resizebox{\textwidth}{35mm}{
\begin{tabular}{cc|cccc|cccc|cccc}
\midrule
\multicolumn{2}{c|}{\multirow{2}{*}{\diagbox{\textbf{Method}}{\textbf{Metric}}}}                                            & \multicolumn{4}{c|}{\textbf{Overall  Performance}}                               & \multicolumn{4}{c|}{\textbf{New Product Performance}}                                                            & \multicolumn{4}{c}{\textbf{L/s Product Performance}}                                                                             \\ \cline{3-14} 
\multicolumn{2}{c|}{}                                                                                                  & \textbf{AUC}    & \textbf{RI}     & \textbf{GAUC}              & \textbf{RI}     & \textbf{AUC}    & \textbf{RI}     & \multicolumn{1}{l}{\textbf{PCOC}} & \multicolumn{1}{l|}{\textbf{Calib-N}} & \textbf{AUC}    & \textbf{RI}                         & \multicolumn{1}{l}{\textbf{PCOC}} & \multicolumn{1}{l}{\textbf{Calib-N}} \\ \midrule
\multicolumn{1}{c|}{\multirow{4}{*}{\begin{tabular}[c]{@{}c@{}}Common \\ DNNs\end{tabular}}}          & DNN            & 0.7358          & -4.57\%         & 0.6557                     & -6.09\%         & 0.7447          & -2.97\%         & \multicolumn{1}{l}{1.1058}        & \multicolumn{1}{l|}{0.01104}          & 0.7359          & -4.61\%                             & 1.0431                            & 0.00183                              \\
\multicolumn{1}{c|}{}                                                                                 & Wide\&Deep~\cite{wideanddeep}     & 0.7412          & -2.39\%         & 0.6624                     & -2.05\%         & 0.7458          & -2.54\%         & \multicolumn{1}{l}{1.2826}        & \multicolumn{1}{l|}{0.08028}          & 0.7414          & -2.39\%                             & 1.1840                            & 0.03377                              \\
\multicolumn{1}{c|}{}                                                                                 & DeepFM~\cite{deepfm}         & 0.7449          & -0.89\%         & 0.6646                     & -0.72\%         & 0.7504          & -0.71\%         & \multicolumn{1}{l}{1.0859}        & \multicolumn{1}{l|}{0.00735}          & 0.7451          & -0.89\%                             & 1.0484                            & 0.00234                              \\
\multicolumn{1}{c|}{}                                                                                 & DIN~\cite{din}            & \underline{0.7471}    & 0.00\%          & \underline{0.6658}                     & 0.00\%          & \underline{0.7522}          & 0.00\%          & \multicolumn{1}{l}{1.1037}        & \multicolumn{1}{l|}{0.01057}          & \underline{0.7473}          & 0.00\%                              & 1.0558                            & 0.00305                              \\ \midrule
\multicolumn{1}{c|}{\multirow{3}{*}{\begin{tabular}[c]{@{}c@{}}Meta-Learning \\ Method\end{tabular}}} & GroupID        & 0.7475          & 0.16\%          & 0.6665                     & 0.42\%          & 0.7526          & 0.16\%          & 1.1617                            & 0.02617                               & 0.7478          & 0.20\%                              & 1.1001                            & 0.00997                              \\
\multicolumn{1}{c|}{}                                                                                 & Meta-E~\cite{meta_emb} & 0.7480          & 0.36\%          & 0.6672                     & 0.84\%          & 0.7533          & 0.44\%          & 1.1184                            & 0.01424                               & 0.7481          & 0.32\%                              & 1.0502                            & 0.00258                              \\
\multicolumn{1}{c|}{}                                                                                 & MWUF~\cite{mwuf}   & 0.7479          & 0.32\%          & 0.6669                     & 0.66\%          & 0.7534          & 0.48\%          & 1.1109                            & 0.01268                               & 0.7481          & 0.32\%                              & 1.0437                            & 0.00199                              \\ \midrule
\multicolumn{1}{c|}{\multirow{2}{*}{Debias Method}}                                                   & LogQ~\cite{logq}           & 0.7485          & 0.57\%          & 0.6673                     & 0.90\%          & 0.7537          & 0.59\%          & 1.3594                            & 0.12950                               & 0.7484          & 0.44\%                              & 1.2963                            & 0.08762                              \\
\multicolumn{1}{c|}{}                                                                                 & ClassRebalance~\cite{class_re} & 0.7470          & -0.04\%         & 0.6651                     & -0.42\%         & 0.7533          & 0.44\%          & 1.0600                            & 0.00384                               & 0.7477          & 0.16\%                              & 0.9917                            & 0.00116                              \\ \midrule
\multicolumn{1}{c|}{\multirow{3}{*}{Gate Method}}                                                     & GateNet~\cite{gatenet}        & 0.7454          & -0.69\%         & \multicolumn{1}{l}{0.6632} & -1.57\%         & 0.7498          & -0.95\%         & 1.1221                            & 0.01363                               & 0.7457          & -0.65\%                             & 1.0639                            & 0.00407                              \\
\multicolumn{1}{c|}{}                                                                                 & FiBiNet~\cite{senet}          & 0.7464          & -0.28\%         & \multicolumn{1}{l}{0.6642} & -0.97\%         & 0.7510          & -0.48\%         & 1.1269                            & 0.01497                               & 0.7467          & -0.24\%                             & 1.0604                            & 0.00364                              \\
\multicolumn{1}{c|}{}                                                                                 & POSO~\cite{poso}           & 0.7464          & -0.28\%         & 0.6663                     & -0.30\%         & 0.7530          & -2.89\%         & \textbf{1.0081}                            & \textbf{0.00005}                            & 0.7465          & -1.05\%                             & \textbf{0.9995}                            & \textbf{0.00024}                              \\ \midrule
\multicolumn{1}{c|}{\textbf{Ours}}                                                                    & MSNet          & \textbf{0.7497} & \textbf{1.05\%} & \textbf{0.6690}            & \textbf{1.93\%} & \textbf{0.7559} & \textbf{1.47\%} & 1.0558                  & 0.00024                     & \textbf{0.7498} & \multicolumn{1}{r}{\textbf{1.01\%}} & 1.0259                   & 0.00065                     \\ \midrule
\end{tabular}}
\caption{
Model comparison on the production dataset. All the lines calculate RelaImpr by comparing with the performance of DIN on different set of products respectively. The best performance for each metric is bold-faced. The improvements are statistically significant (i.e. two-side t-test with $p < 0.05$) over the original model, where "RI" is short for "RelaImpr", and "L/s" is short for "limited-stock".
}
\label{tab:offline_result}
\end{table*}

\begin{table}[t]
\begin{tabular}{lcccc}
\toprule
\diagbox{\textbf{Method}}{\textbf{Metric}} & \textbf{AUC}                  & \textbf{RelaImpr} & \textbf{GAUC}                 & \textbf{RelaImpr} \\ \hline
Base                         & \underline{0.7471}                        & 0.00\%           & \underline{0.6658}                        & 0.00\%            \\
W/o seq-split              & {0.7484} & 0.53\%            & {0.6671} & 0.78\%            \\
W/o seq-meta            & {0.7481} & 0.40\%            & {0.6669} & 0.66\%            \\
W/o auxiliary loss                 & {0.7489} & 0.73\%            & {0.6679} & 1.26\%            \\
\textbf{MSNet}                        & \textbf{0.7497}                        & \textbf{1.05\%}           & \textbf{0.6690}                        & \textbf{1.93\%}            \\ \toprule
\end{tabular}
\caption{
    Ablation study of different MSNet variants on production datasets, where "w/o" is short for "without".
}
\label{tab:ablation result}
\end{table}

\begin{table}[]
\begin{tabular}{cccc}
\toprule
                     & \textbf{CTR} & \textbf{Clicks} & \textbf{Exposure Ratio (absolute)} \\ \hline
\textbf{Overall}     & +3.56\%      & +3.31\%         & -                            \\
\textbf{L/s Product} & +3.72\%      & +5.20\%         & +1.00\%                      \\
\textbf{M/s Product} & +3.64\%      & +1.44\%         & -1.00\%                      \\ \toprule
\end{tabular}
\caption{
Comparison between production model and MSNet in online A/B tests, where "L/s" is short for "limited-stock", and "M/s" is short for "multiple-stock". We followed industry standards and best practices for online A/B testing, ensuring reliability and validity through randomized assignment and control of external factors that could bias the results.
}
\label{tab:online result}
\end{table}

In this section, we conduct extensive experiments on the offline
dataset and online A/B testing to evaluate the performance of the proposed method. The following research questions (RQs) are addressed:
\begin{itemize}
    \item \textbf{RQ1}: Does your proposed method outperform state-of-the-art recommender? (Section 5.2)
    \item \textbf{RQ2}: Does your proposed method address the issue of limited-stock product in CTR prediction? (Section 5.2)
    \item \textbf{RQ3}: How do critical components affect the performance of your proposed method? (Section 5.3)
    \item \textbf{RQ4}: Does your proposed method work in real large-scale online recommendation scenarios? (Section 5.4)
\end{itemize}
\subsection{Experiment Setup}

\subsubsection{\textbf{Xianyu Dataset}}
We collect click traffic logs of 8 days from Alibaba’s Xianyu Recommendation Platform\footnote{To the best of our knowledge, there are no suitable public data-set for CTR prediction under the isolate item problem.} to build the production data-set with 10 billion samples (1.5 billion sample per day), with numbers of items and people, 209 features (e.g., user and item features). 
The samples in the first 7 days and 8th day are employed for training and testing respectively, and we randomly partition the testing set into 10 parts and report average evaluation results.

\subsubsection{\textbf{Baselines}}
We compare the proposed MSNet with four types of baselines: the first type are common DNN-based methods, including:
\begin{itemize}
    \item \textbf{DNN}: a method proposed by Youtube, and has been been implemented in various industrial applications.
    \item \textbf{Wide \& Deep} \cite{wideanddeep}:  a method proposed by Google that consists of a wide model and a deep model to and has gained significant adoption in various applications.
    \item \textbf{DeepFM} \cite{deepfm}: a method proposed by Huawei Noah lab that that combines deep learning and factorization machine techniques.
    \item \textbf{DIN} \cite{din}: our base model, as described in Section 4, which uses attention mechanism to  aggregate historical behaviors sequence information based on the similarity between historical behaviors item and the target item.
\end{itemize}
The second type are meta learning methods
\begin{itemize}
    \item \textbf{GroupID}~\cite{groupid} a meta-learning approach proposed by Airbnb to generate new item embeddings by calculating the average of similar item embeddings. 
    \item \textbf{Meta-E}~\cite{meta_emb}: a meta-learning approach to address the cold-start problem by generating ID embeddings for cold items using other available features.
    \item \textbf{MWUF}~\cite{mwuf}: a meta-learning method to address the cold-stard problem, which consists of two parts. The Scaling Network can produce a scaling function to transform the cold ID embedding into warm feature space, while the Shifting Network is able to produce a shifting function that can produce stable embedding from the noisy embedding. 
\end{itemize}
The third type are some gate network method, because our method only boosts the embedding part, we only implement the embedding gate for following methods.
\begin{itemize}
    \item \textbf{GateNet} \cite{gatenet}: a method proposed by Baidu, using gate layer to generate weight for the embedding layer and the hidden layer.
    \item \textbf{FiBiNet} \cite{senet}: a method proposed by Sinanet, filtering the information of embedding based on their importance using the SENet module.
    \item \textbf{POSO} \cite{poso}: a method proposed by KuaiShou, a short-video application. We implement it as described in the paper with personal MLP and personal MHA.
\end{itemize}
As for the fourth type of baselines, we implement two label de-bias approaches
\begin{itemize}
    \item \textbf{Log-Q} \cite{logq}: it introduces an item frequency-based factor to modify the softmax logit within the loss function. This modification allows for differential weighting of head and tail items during the learning process.
    \item \textbf{ClassRebalance} \cite{class_re}: a method to address class imbalance, which computes the effective number of samples for each imbalanced class and adopts it as a re-weighting factor within the loss function.
\end{itemize}

\subsubsection{\textbf{Evaluation Metric}}
Following previous works \cite{din,dien,star,clid}, we adopt some widely used metrics, AUC, GAUC and RelaImpr, to evaluate the performance of our proposed method. 
The calculation of AUC, GAUC, RelaImpr can be expressed as follow:
\begin{equation}
\mathrm{AUC}=\frac{1}{|P||N|} \Sigma_{p \in P} \Sigma_{n \in N} I(\Theta(p)>\Theta(n))
\end{equation}
where $P$ and $N$ denote positive sample set and negative sample set, respectively. $\Theta$ is the estimator function and $I$ is the indicator function.

\begin{equation}
\mathrm{GAUC}=\frac{\sum_{i=1}^n \# \text { impression }_i \times \mathrm{AUC}_i}{\sum_{i=1}^n \text { \#impression }_i},
\end{equation}
where the AUC is first calculated within samples of each user, and averaged w.r.t sample count, where $n$ is the number of users, \#impression $i$ and $\mathrm{AUC}_i$ are the number of impressions and AUC corresponding to the $i$-th user.

\begin{equation}
\text { RelaImpr }=\left(\frac{\text { AUC }(\text { measured model })-0.5}{\text { AUC }(\text { base model })-0.5}-1\right) \times 100 \%
\end{equation}
where RelaImpr metric to measure relative improvement over models and  0.5 stands for the AUC of a random guesser.

In order to mitigate the interference caused by the distribution of the prediction dataset, we randomly partition the users into ten groups. We calculate the AUC for each group separately and then compute the average and standard deviation of these ten AUC metrics, which are represented as avg(AUC) and std(AUC).

Furthermore, to assess the accuracy of the absolute values of the model scores, we utilize the Cal-N metric to represent the calibrated error between the model scores and the predicted probabilities of the label in the prediction dataset, which is calculated as follows: 

\begin{equation}
\mathrm{Cal}\text{-}N=\sqrt{\frac{\sum_{i=1}^{N}error_i^2}{N}}
\end{equation}

\begin{equation}
error_i = \left\{
\begin{array}{ll}
\mathrm{PCOC}_i - 1, \mathrm{PCOC}_i \geq 1 \\
\frac{1}{\mathrm{PCOC}_i} - 1, \mathrm{PCOC}_i < 1 \\
\end{array}
\right.
\end{equation}

\begin{equation}
\mathrm{PCOC}_i=\frac{\sum_{j\in D_i} \hat{p}_j}{\sum_{j\in D_i} p_i}
\end{equation}

The PCOC (Predict Click Over Click) metric is calculated over each sub-dataset split by N partitions and then the error of each sub-dataset is calculated by~\cite{cvr1}. Finally, the calibration-N metric can be obtained by calculating the standard deviation of N errors. A smaller Cal-N indicates higher accuracy in the absolute values of the model scores.

In general, for AUC and GAUC, the higher the value, the better the performance. For PCOC, the closer to 1, the better the performance. For Calib-N, the smaller the value, the better the performance.

\subsubsection{\textbf{Implementation Details.}}
We implement the DNN part of DeepFM, Wide \& Deep, DNN, DIN all the same architecture, i.e., a three-layer MLP Network with 512, 256 and 128 hidden units. For all attention layers in above models, the number of hidden units are all set to 128. AdagradDecay optimizer is adopted in all the methods, the learning
rate of 1e-4 is set. Batch size are set to 4096 for all DNNs. And the histoty click sequence are collected with last 30days and max length are 50.

% copy from GIFT, much too similary? 

\subsection{RQ1 \& RQ2: Overall Comparison with Baselines}
We conduct an evaluation of different approaches on the production dataset for CTR prediction tasks and the experimental results are presented in Table~\ref{tab:offline_result}. 
Based on the information provided in the table, we have the following observations:
\begin{itemize}
% ob1: 我们的方法在整体物品的指标上 outperform all other method， 相比于base model DIN来说，我们的auc提升了1.05\%, GAUC提升了 1.93%, specifically，xxx，which answers our RQ1

% ob2: 我们的方法在部分物品的auc指标也是最好的，除了在细节精度指标一些指标，缓解了我们场景下浅库存商品问题，specifically，相比于base model DIN来说，,新品auc提升了1.47\%, pcoc从1.1037降低到1.0558， calib-n从0.01057降低到0.00274也有降低，对于C品来说，auc提升了1.01\%, pcoc从1.1037降低到，calib-n从0.01057降低到0.00274也有降低， 1.0558，specifically which answers our RQ2

% ob3: 其他方法有的能提升，有的不能提升。对于meta learning的方法来说，MELU，MWUF的提升比GroupID要好，原因可能是groupid通过同类目下热门商品的item id来代替新品的item id，可能类目不是那么的精细，商品之间和商品之间不是那么的相似，item id不能直接适用，相反，MELU和MWUF通过商品别的信息来生成item id信息，会带来一定的提升，但是没有考虑到库存和序列，所以提升不如我们。对于gate方法来说，单纯的增加门控网络的参数，会让原本就已经倾斜的dnn网络更加倾斜，从而无法带来提升。用样debias通过修改loss也带了了提升

    \item Our method outperformed all other methods in terms of overall item metrics. Specifically, compared to the base model DIN, our approach achieved a 1.05\% improvement on AUC and a 1.93\% increase on GAUC, which answers our research question RQ1.~\footnote{Note that the 0.1\% AUC gain is already considerable in large-scale industrial recommendation system \cite{chang2023pepnet}, especially when we didn't add additional information}
    
    \item Our method also achieved the highest AUC metric for certain items, except for some precision-related metrics. It effectively mitigated the issues related to limited-stock items in our scenario. Specifically, compared to the base model DIN, our approach resulted in a 1.47\% improvement in AUC for new items, a decrease in pcoc from 1.1037 to 1.0558, and a decrease in calib-n from 0.01057 to 0.00274. For type C items, our method showed a 1.01\% improvement in AUC, a decrease in pcoc from 1.1037 to 1.0558, and a decrease in calib-n from 0.01057 to 0.00274. These results address our research question RQ2.

    \item Other methods were not able to effectively address our problem, although some methods showed improvements compared to the base model. In the context of meta learning methods, MELU and MWUF outperformed GroupID. This could be because GroupID replaces the item ID of new items with the ID of popular items in the same category, which may not accurately capture the fine-grained similarities between different items. On the other hand, MELU and MWUF generate item ID information based on other characteristics of the items, leading to better performance. However, these methods do not consider factors such as inventory and sequence information, resulting in less significant improvements compared to our method. Similarly, the debias method showed improvements by modifying the loss function.As for the gate method, simply increasing the parameters of the gate control network can further bias the already biased DNN network, making it unable to bring any improvement. In the future, when we deploy our system, we will also consider integrating algorithms that can provide benefits.
\end{itemize}

\subsection{RQ3: Ablation Study}
To further validate the effectiveness of the sub-modules proposed in the MSNet model, we compare the offline performance of models without the sequence split module, without the seq meta learning module, without auxiliary loss, without all modules, and the complete model, as shown in Table~\ref{tab:ablation result}.
% 证明了， 序列拆分的必要性， 序列信号增强的必要性， 以及 辅助更新 seq item id 的必要性
The results clearly demonstrate the effectiveness of our proposed MSNet, highlighting the necessity of sequence splitting, enhancing sequence signals, through sequence meta learning, and addtional auxiliary loss in updating sequence item IDs, respectively.

\subsection{RQ4: Online A/B test}
% We conduct online A/B testing in the recommendation system of Xianyu to evaluate its effectiveness. 
% The compared baseline is a highly-optimized model with the form of DIN. 
% To get a stable conclusion, we observe the online experiments for 7 days and we use CTR, Clicks and exposure ratio to evaluate the online effectiveness. As we can see in Table~\ref{tab:online result}, MSNet achieved a 3.56\% improvement in CTR and a 3.31\% improvement in Clicks compared to base model,  which can bring huge benefits to the e-commerce platform. It is notable that we improve more on the performance of limited-stock product with a 3.72\% improvement in CTR, a 5.20\% improvement in Clicks and a 1.0\% absolute gain in exposure ratio.

% We conducted online A/B testing in the recommendation system of Xianyu to evaluate its effectiveness. The baseline model we compared against was a highly-optimized model with the form of DIN. To ensure a stable conclusion, we observed the online experiments for a duration of 10 days. We used CTR (Click-Through Rate), Clicks, and exposure ratio as metrics to evaluate the online effectiveness. The results are shown in Table~\ref{tab:online result}.
In Xianyu's recommendation system, we compared a new model's effectiveness against a highly-optimized baseline (DIN architecture) through an online A/B test. Users were randomly assigned to control and experimental groups, with device IDs uniformly distributed via MD5 hashing for impartial partitioning. The 10-day experiment period provided statistically validated, reliable results.We used CTR (Click-Through Rate), Clicks, and exposure ratio as metrics to evaluate the online effectiveness. The results are shown in Table~\ref{tab:online result}.

% The MSNet model achieved a 3.56\% improvement in CTR and a 3.31\% improvement in Clicks compared to the base model. These improvements can bring significant benefits to the e-commerce platform. It is worth mentioning that we observed even greater improvements in the performance of limited-stock products. We saw a 3.72\% improvement in CTR, a 5.20\% improvement in Clicks, and a 1.0\% absolute gain in exposure ratio for limited-stock products.

The MSNet model yielded a notable improvement across key metrics compared to the base model, with a 3.56\% increase in CTR and a 3.31\% increase in Clicks. Particularly significant were the enhancements observed in limited-stock products, with a 3.72\% increase in CTR, a 5.20\% increase in Clicks, and a 1.0\% absolute gain in exposure ratio. 

Regarding time complexity, MSNet architecture allows parallel processing with minimal computational cost. The new introduced meta-DNN, utilizing a small MLP network, requires limited resources. In our online environment, MSNet's implementation increases response time by less than 5\%, acceptable for our recommendation system.

\section{Conclusion}

In this study, we present the Meta-Split Network (MSNet) for enhancing click-through rate (CTR) predictions in C2C e-commerce, especially for limited-stock products. Traditional models falter with limited-stock items due to sparse data affecting item embeddings. MSNet tackles this by segmenting user history based on stock levels and applying a unique meta-learning approach for low-stock items. This method, supported by meta scaling and shifting networks, optimizes performance. Additionally, an auxiliary loss component updates item embeddings beyond consumption. Demonstrated effective through experiments and online A/B testing, MSNet emerges as a pioneering solution for C2C limited-stock recommendations. It offers valuable insights and approaches for similar challenges in e-commerce recommendation systems.

% %% consistent spelling of the heading.
% \begin{acks}
% \end{acks}

\bibliographystyle{ACM-Reference-Format}
\balance
\bibliography{main}

% \bibliographystyle
% \balance
% \bibliography


\end{document}